\documentclass[
  reprint,              
  aps,                  
  pra,                  
  superscriptaddress,   
  amsmath,amssymb,      
  floatfix,             
  longbibliography      
]{revtex4-2}

\usepackage{graphicx}   
\usepackage{dcolumn}    
\usepackage{bm}         
\usepackage[usenames,dvipsnames]{xcolor}
\usepackage{hyperref}   
\hypersetup{
  colorlinks=true,
  linkcolor=blue,
  citecolor=blue,
  urlcolor=blue,
  pdfauthor={Your Name},
  pdftitle={Title of the Manuscript}
}

\usepackage[T1]{fontenc}
\usepackage[utf8]{inputenc}
\usepackage{microtype}  
\usepackage{siunitx}    
\usepackage{xcolor} 

\newif\ifshowedits
\showeditsfalse 

\newcommand{\blue}[1]{%
  \ifshowedits
    \textcolor{blue}{#1}%
  \else
    {#1}
  \fi
}

\begin{document}

\preprint{APS/123-QED}

\title{Topological transitions controlled by the interaction range}
\author{Vlad Simonyan}
\email{vlad.simonyan@metalab.ifmo.ru}
\affiliation{School of Physics and Engineering, ITMO University, Saint  Petersburg 197101, Russia}

\author{Maxim A. Gorlach}
\affiliation{School of Physics and Engineering, ITMO University, Saint  Petersburg 197101, Russia}


\begin{abstract}
We study a one-dimensional topological model featuring a Su-Schrieffer-Heeger type pattern of nearest-neighbor couplings  in combination with the longer-range interactions exponentially decaying with the distance. We demonstrate that even relatively weak long-range couplings can trigger the topological transition if their range is large enough. This provides an additional facet in the control of topological phases.
\end{abstract}

\maketitle


\section{\label{sec:intro}Introduction}

Topological systems have attracted much attention due to the immunity of their modes to backscattering and resilience of the mode energies to certain disorder types~\cite{Ozawa_2019,Lu2014,Khanikaev2017}. By now, the potential of such systems is well appreciated for a plethora of material platforms including condensed matter~\cite{Hasan2010,Xiao2010,zhang2019catalogue}, photonics~\cite{Ozawa_2019,Lu2014,Khanikaev2017,Rechtsman2013,Hafezi2013,Khanikaev2012}, polaritonics~\cite{Karzig2015,Klembt2018}, cold atoms~\cite{Cooper2019,Zhu2018} and others.


The theoretical description of such systems often builds on the tight-binding Hamiltonians incorporating only nearest-neighbor interactions~\cite{Shen,Asboth}, the one-dimensional (1D) Su-Schrieffer-Heeger model (SSH)~\cite{Su1979,Heeger1988,Shen,Asboth} being the simplest and well studied example. This assumption keeps the analysis simple while capturing the underlying physics in many instances.

Adding to this picture longer-range interactions can change the topological phase diagram not only quantitatively, but also qualitatively~\cite{PerezGonzalez2018}. Typically, the condition for such change is the sufficient strength of the long-range interaction, which should become comparable to the nearest-neighbor couplings. This condition is readily met in the one-dimensional arrays of dielectric meta-atoms, where the dispersion of the Bloch modes is strongly modified compared to the tight-binding scenario~\cite{He2022}. The similar phenomena take place for trapped ion systems~\cite{nevado2017topological} and even optical waveguides~\cite{Schulz:22}, becoming even more pronounced in the 2D case owing to the larger number of the neighboring sites~\cite{Li2019,Olekhno2022}.


However, the impact of weak, but long-radius interaction in the context of topological phases is understood much less. In fact, most of extended versions of topological models truncate the hopping range to the finite set of neighboring sites and hence do not explore the role of the interaction length. \blue{As a result, the main tuning parameter is the overall strength of additional couplings, while the interaction length is kept fixed or effectively truncated.} To fill this gap, we study here a one-dimensional topological model involving long-range couplings with a smooth dependence on the distance. Besides the strength of the long-range interaction $J$ we examine the role of another parameter $\lambda$ quantifying the interaction range. Counterintuitively, we find out that even very weak long-range couplings enable topological transitions if their radius is large enough.

\section{\label{sec:methods}Model}

\begin{figure}[b]
  \includegraphics[width=1\linewidth]{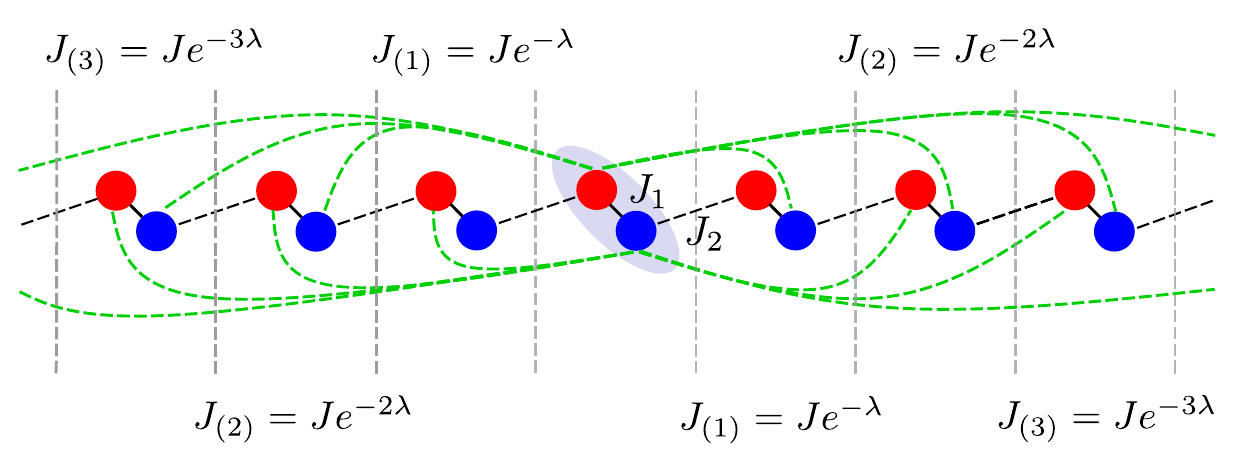}
    \caption{Schematic illustration of the Su–Schrieffer–Heeger model (SSH) extended by the long-range couplings. The coupling amplitudes decay exponentially with the distance as \( J_{(n)} = Je^{-n\lambda} \), where \( \lambda \) controls the interaction range and $n$ is the difference between the unit cells coordinates. Red and blue sites depict the two sublattices.}
  \label{fig:chain}
\end{figure}

As a specific example, we investigate an extended  Su-Schrieffer-Heeger model with additional long-range couplings. These extra links connect sites from the different sublattices as illustrated in Fig.~\ref{fig:chain}. This specific choice of the long-range couplings preserves the chiral symmetry of the model ensuring that the operator $\hat\Gamma=\sigma_z$ anticommutes with the Bloch Hamiltonian:
$\{ \Gamma, \hat{H}(\mathbf{k}) \} = 0$. 
Equivalently,
\begin{equation}
\Gamma \, \hat{H}(\mathbf{k}) \, \Gamma^{-1} = - \hat{H}(\mathbf{k}),
\label{eq:ChiralSym}
\end{equation}
which guarantees that the spectrum is symmetric with respect to zero energy and enables zero-energy edge-localized modes.

In our treatment, all lattice sites have the same on-site energy (i.e. resonance frequency) which is set to zero for convenience. The nearest-neighbor couplings are either $J_1$ or $J_2$, as in the canonical SSH lattice. In addition to that, we include long-range couplings that connect the sites of the different sublattices: $A_n$ to $B_m$ with $m\neq n$. We assume that the coupling  depends only on the difference of the unit cell coordinates $s = |n-m|$ and decreases exponentially with the distance, $J_{(s)} = J e^{-\lambda s}$, where $J$ sets the overall scale of long-range hopping and $\lambda$ is the inverse of characteristic interaction length.  Thus, the overall model combines the alternating coupling pattern of the SSH lattice with spatially extended interactions.

The real-space Hamiltonian of our model can be written compactly as:
%
\begin{align}
\hat{H} &= 
\sum_n \left( 
    J_1 \hat{a}^\dagger_n \hat{b}_n 
  + J_2 \hat{a}^\dagger_{n+1} \hat{b}_n
\right) +  \nonumber \\
&\quad + \sum_{n} \left( 
    \sum_{s=1}^{\infty} J_{(s)} \hat{a}^\dagger_n \hat{b}_{n+s} 
  + \sum_{s=2}^{\infty} J_{(s)} \hat{a}^\dagger_{n+s} \hat{b}_n
\right) 
+ \text{H.c.}\:,
  \label{eq:RSHam}
\end{align}
where $\hat{a}_n^\dagger,\hat{b}_n^\dagger$ and $\hat{a}_n,\hat{b}_n$ denote the creation and annihilation operators at the $n^{\text{th}}$ site of $A$ or $B$ sublattices, respectively. Note also that the limit $\lambda\rightarrow\infty$ corresponds to the canonical SSH model.

\section{\label{sec:res}RESULTS}

 We perform a Fourier transform of the Hamiltonian, which leads to its Bloch representation $H(k)$:
\begin{equation}
\hat{H} = 
\sum_{\mathbf{k}}
\begin{pmatrix}
\hat{a}_{\mathbf{k}}^{\dagger} & \hat{b}_{\mathbf{k}}^{\dagger}
\end{pmatrix}
\underbrace{\begin{pmatrix}
0 & h(k) \\
h^*(k) & 0
\end{pmatrix}}_{H(\mathbf{k})}
\begin{pmatrix}
\hat{a}_{\mathbf{k}} \\
\hat{b}_{\mathbf{k}}
\end{pmatrix}\:.
\label{eq:BlochHam}
\end{equation}

In this form, the nearest-neighbor and long-range couplings contribute only to the off-diagonal block $h(k)$ of the Bloch Hamiltonian maintaining the chiral symmetry. As long-range interactions decay exponentially with the distance, the resulting sums of couplings are readily evaluated analytically. This yields a closed-form expression for the off-diagonal element $h(k)$:
\begin{equation}
\begin{aligned}
h(k) &= J_1+(J_2-J_{(1)})e^{-ik}+\sum_{s=1}^{\infty}\left( J_{(s)}e^{-iks}+ J^*_{(s)}e^{iks}\right) =  \\
     &= J_1+(J_2-Je^{-\lambda})e^{-ik}+J\frac{\cos k - e^{\lambda}}{\cosh \lambda - \cos k}\:.
\end{aligned}
\label{eq:hk}
\end{equation}
%

Such an off-diagonal structure of the Hamiltonian naturally allows us to define a topological invariant in the form of the winding number \cite{topological_ryu_2002}:
\begin{equation}
w = \frac{1}{2\pi i} \int_{-\pi}^{\pi} d k  \frac{d}{dk} \ln h(k).
  \label{eq:wnum}
\end{equation}

\noindent The winding number counts how many times $h(k)$ winds around the coordinate origin in the complex plane as $k$ spans the Brillouin zone. Different integer values of $w$ label distinct topological phases.

\begin{figure}
  \includegraphics[width=1\linewidth]{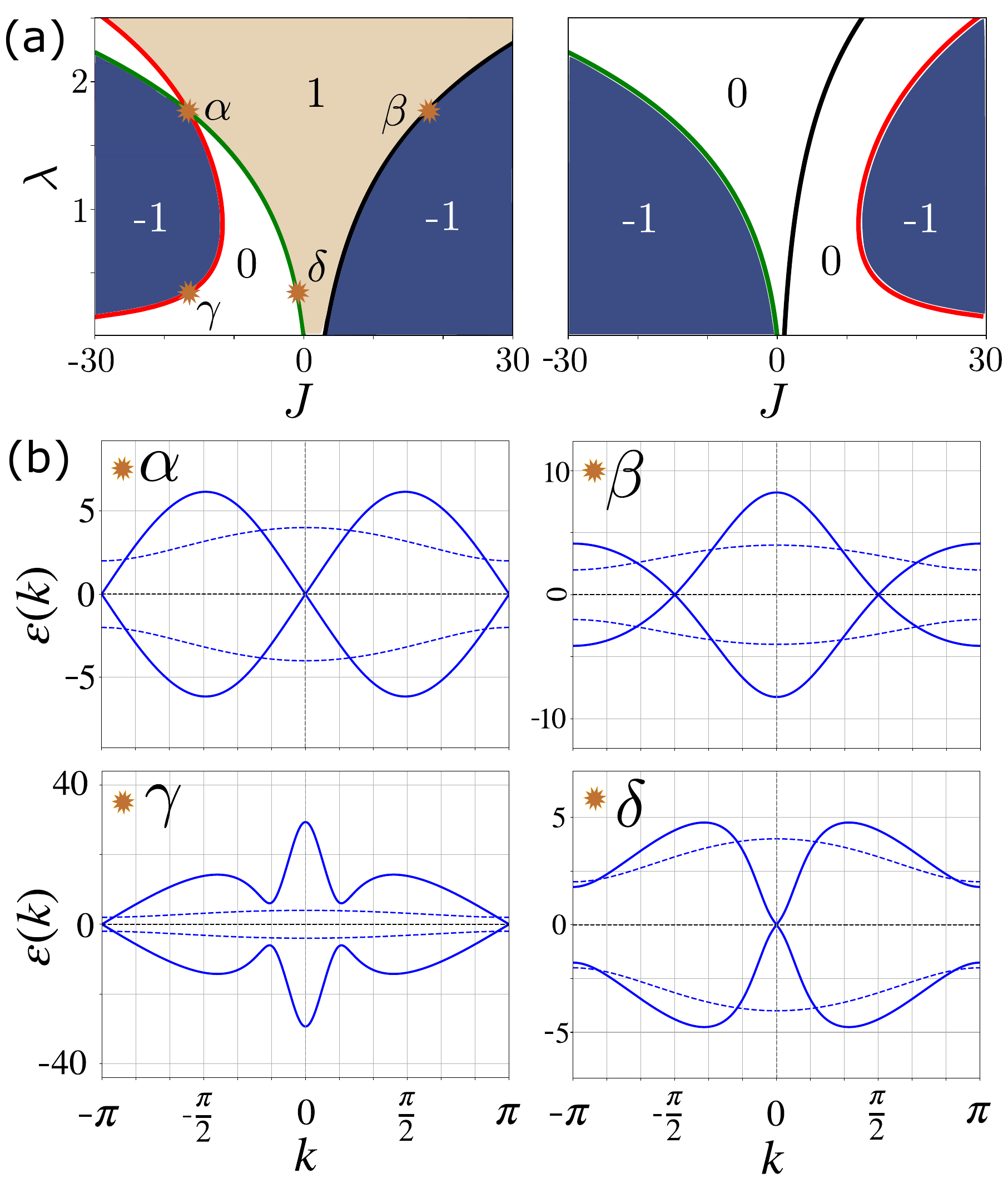}
    \caption{ (a) Winding number phase diagrams in the \( (J, \lambda) \) parameter space for the different  dimerizations of the lattice: 
        (left) \( J_1=1 \), \( J_2=3 \); (right) \( J_1 = 3 \), \( J_2 = 1 \). $J_1$ link is inside the unit cell. Distinct topological phases characterized by the different winding numbers \( w \in \{ -1, 0, 1 \} \) are color-coded. 
        The phase boundaries corresponding to band gap closing are marked by red, blue and black curves.
        (b) Band diagrams calculated for the parameters indicated by the points $\alpha,\beta,\gamma$ and $\delta$ in panel (a). Dashed lines showing the respective dispersion for the canonical SSH model are provided for reference.
    }
    \label{fig:phases}
\end{figure}

The change in the winding number \(w\) occurs only for such parameters when the curve \(h(k)\) in the complex \((h_x(k),h_y(k))\) plane touches the origin \((0,0)\), with \(h_x(k)=\text{Re}\, h(k)\) and \(h_y(k)=\text{Im}\, h(k)\). This yields the analytical expressions for the phase transition curves in the \((J,\lambda)\) parameter space:
\begin{equation}
\left[
\begin{aligned}
J &= \frac{(J_1 - J_2)e^{\lambda}}{\tanh(\lambda/2)}, \\
J &= -(J_1 + J_2)e^{\lambda}\tanh(\lambda/2), \\
J &= J_2 e^{\lambda}
\end{aligned}
\right.
  \label{eq:lines}
\end{equation}

Equations~(\ref{eq:lines}) match precisely the numerically computed phase boundaries shown in Fig.~\ref{fig:phases}(a). The phase diagram features the regions with the winding numbers $w=0$ and $w=\pm 1$.  \blue{Note that the phases with $w=1$ and $w=-1$ are related by the time-reversal transformation which leaves invariant the zero-energy edge mode. Hence, the sign of the winding number is not imprinted in the properties of the edge state.} Fixing the interaction range $\lambda$ and gradually increasing the absolute value of the long-range coupling, we observe the change in the winding number, i.e. the topological transition. This fully aligns with the previous studies indicating that relatively strong long-range couplings can induce the topological transition.

Interestingly, the change in $\lambda$ for a fixed $J$ induces the topological transition as well. In fact, even  vanishingly small long-range interaction $J<0$ can trigger the topological transition if its range $\lambda^{-1}$ is large enough. The threshold value of $\lambda$ favoring such a transition is estimated from Eq.~\eqref{eq:lines} as
\begin{equation}
    \lambda_{\text{cr}}\approx-\frac{2J}{J_1+J_2}\:.
\end{equation}
%

 \blue{ Physically, this means that many weak long-range links contribute coherently to the $k=0$ Bloch state, and their total effect is equivalent to an effective coupling of the order of $2J/\lambda$. When this cumulative contribution compensates the sum of the nearest-neighbour couplings $J_1+J_2$, the off-diagonal matrix element vanishes at $k=0$ and the band gap closes, thus producing the topological transition. }

In addition, long-range interactions enable yet another topological sector with $w=-1$, making the phase diagram richer compared to the standard SSH. From a topological perspective, the phases with $w=1$ and $w=-1$ are equivalent since in both cases the Zak phase~\cite{Zak} equals $\pi$. However, the transition between $w=-1$ and $w=1$ necessarily involves a gap closing, which defines the characteristic curves shown in Figs.\ref{fig:phases}(a,b) by the black lines. 

The calculated band structures for several representative parameter sets shown by the points $\alpha$, $\beta$, $\gamma$, and $\delta$ in Fig.~\ref{fig:phases}(a) are depicted in Fig.~\ref{fig:phases}(b). Contrary to the canonical SSH model, these band structures feature closing and reopening of gaps not only at the center and edges of the Brillouin zone, but also in the intermediate points highlighting an important role of the long-range couplings. \blue{In some cases the bandgap closes at an intermediate momentum $k_\star \approx \pi/2$ [see black curves in Fig.~\ref{fig:phases}(a) and the dispersion in Fig.~\ref{fig:phases}(b), panel $\beta$]. At this point, the trajectory of $h(k)$ in the complex plane collapses into a straight segment passing through the origin and then bends to the opposite side when the parameters are tuned across the line at the phase diagram. Depending on whether this trajectory encircles the origin or not, such gap closing may either change the sign of the winding number [$w=1 \leftrightarrow w=-1$ in the left panel of Fig.~\ref{fig:phases}(a)] or leave $w=0$ unchanged, as in the right panel.
}

\begin{figure}
  \centering
  \includegraphics[width=0.8\linewidth]{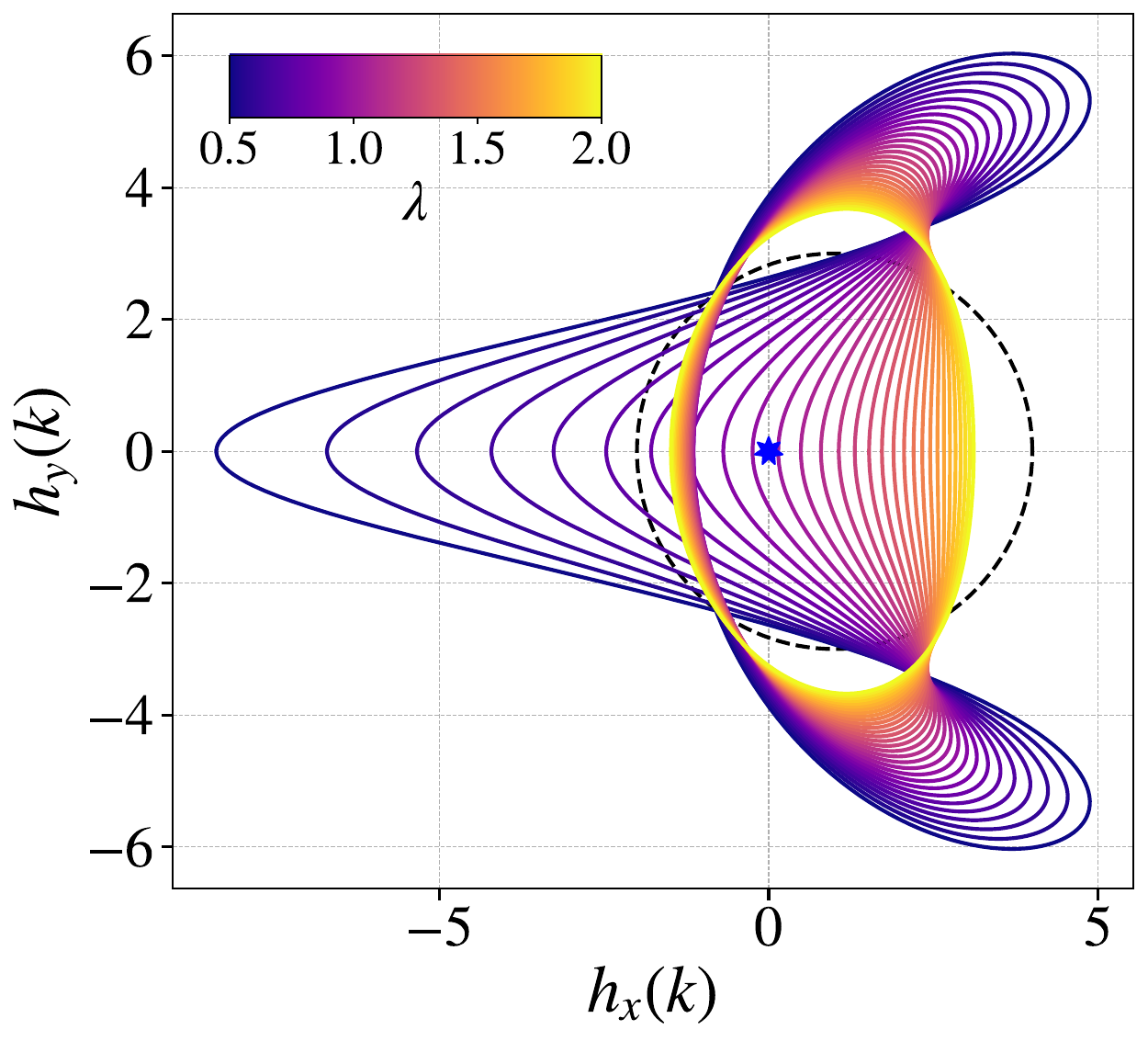}
  \caption{Evolution of the winding curve as a function of $\lambda$. The black dashed circle corresponds to the case without long-range couplings, which is the canonical SSH model. The colored curves show the trajectories of $h(k)$ for $J_1=1, J_2=3,J=-5$ and different values of $\lambda$ which are color-coded. The blue star marks the origin $h(k)=0$.}
  \label{fig:winding_diagrams}
\end{figure}

To illustrate the topological transition driven by the change of $\lambda$, Fig.~\ref{fig:winding_diagrams} shows the evolution of the winding curve as the parameter \(\lambda\) is varied for the fixed $J$. Larger $\lambda$ result in an elliptical curve very much resembling one in the canonical SSH. However, as $\lambda$ decreases, i.e. the interaction range grows, the winding curve rapidly deforms, crosses the origin, thereby inducing a topological phase transition. The computed evolution of the winding curve also explains why our model allows $w=0$ and $w=\pm 1$, but prohibits higher winding numbers.


The presence of nonzero winding numbers implies, via the bulk-boundary correspondence~\cite{topological_ryu_2002,Asboth,Shen}, that topologically protected zero-energy edge states must exist in the corresponding phases.

To examine the onset of the localized states, we perform numerical diagonalization of the Hamiltonian matrix for a finite system with the open boundary conditions. To quantify the localization of the zero-energy modes, we compute the inverse participation ratio (IPR)~\cite{THOULESS197493}
\begin{equation}
\text{IPR} = \sum_{n} |\psi_n|^4\:,
\label{eq:ipr}
\end{equation}

\noindent where $\psi_n$ is the wave function amplitude at site $n$ and the wave function is normalized as $ \sum_{n} |\psi_n|^2=1$. The IPR provides a measure of localization: for an extended state in 1D the IPR scales as $1/N$, where $N$ is the number of the lattice sites, while for exponentially localized states IPR does not depend on $N$ if the array is long enough.

 \blue{We emphasize that although the additional hoppings decay exponentially with distance, the model remains effectively local: each site is strongly coupled only to a finite number of the neighboring sites, as set by the interaction length $\lambda^{-1}$. This ensures that the bulk-boundary correspondence is maintained.}


\begin{figure*}[t]
  \includegraphics[width=0.8\linewidth]{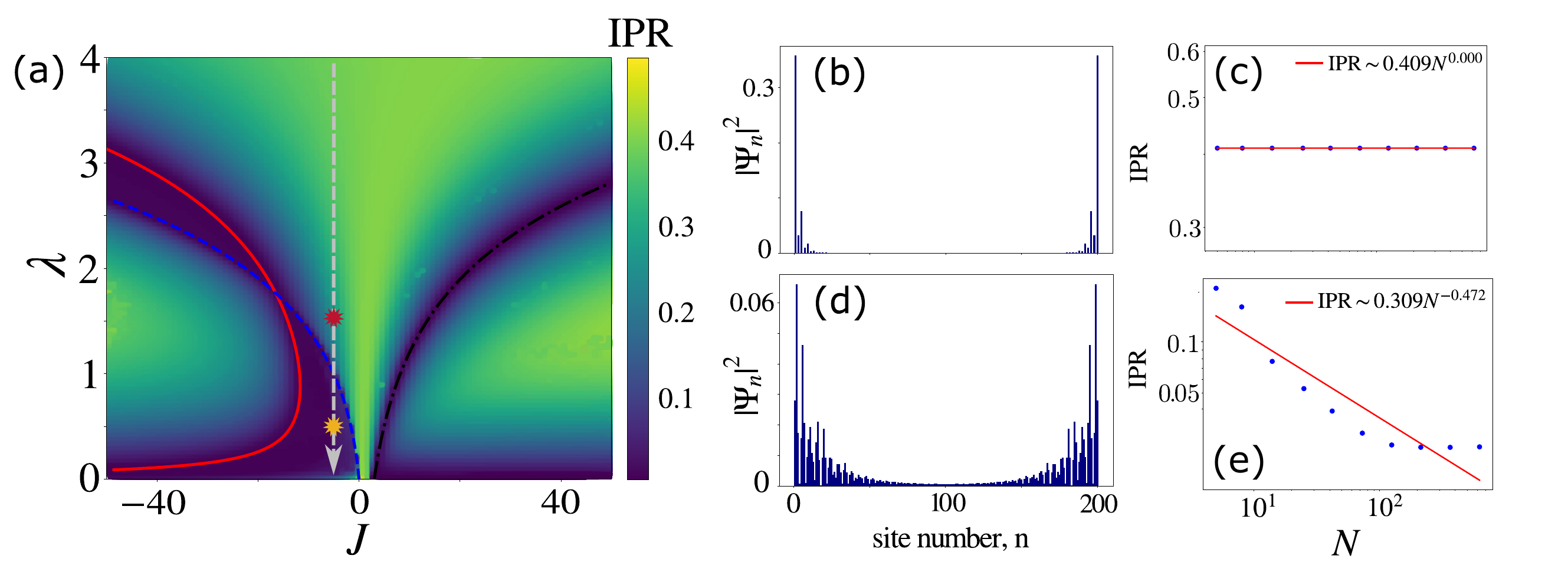}
    \caption{(a) The map of inverse participation ratio (IPR) for the zero-energy mode in an extended SSH model
    with the nearest-neighbor couplings \( J_1 = 1 \), \( J_2 = 3 \) and system size \( N = 100 \). 
    The red, blue and black curves indicate bandgap closing points, the silver line illustrates the phase transition driven by the interaction length.
    (b) and (c) are the spatial profile of the zero energy mode  \blue{with energy $\varepsilon\approx0$} and IPR dependence on the array length $N$ for the parameters shown by the red star in panel a: $\lambda = 1.5, J = -5$. (d,e) present the same data for another set of parameters, gold star in panel a, $\lambda = 0.5, J = -5$,  \blue{$\varepsilon \approx 1.23$ which is the closest to the zero energy}.
    }
    \label{fig:ipr}
\end{figure*}

Figure~\ref{fig:ipr}(a) presents the computed IPR map, which closely matches the phase diagram shown in Fig.~\ref{fig:phases}(a), indicating full consistency between the analysis of a periodic system and direct numerical diagonalization for a finite chain. 
To illustrate this further, we select two representative points marked by stars on the phase diagram [Fig.~\ref{fig:ipr} (a)]. One of the points lies in the topological phase and another one is in the trivial phase.  \blue{In the topological phase, the array with open boundary conditions supports two zero-energy edge modes, one at each boundary, which weakly hybridize in a finite system so that each eigenmode has finite weight at both edges, as seen in Fig.~4(b).
} At first glance, the state in the trivial phase appears to be localized [Fig.~\ref{fig:ipr} (d)]. However, a careful analysis of the IPR scaling with the array length $N$ [Fig.~\ref{fig:ipr} (e)] reveals an extended nature of the mode: the IPR goes down as the array length $N$ increases. This confirms that truly localized edge states persist only in the topological sector.  \blue{We note that, apart from modifying the phase diagram, the exponentially decaying long-range couplings do not qualitatively change the spatial structure of the zero-energy edge states, which remain exponentially localized at the boundaries with a localization length set by the renormalized bulk gap.
}

\section{Conclusions}
In summary, our study demonstrates that the impact of the long-range interactions depends not only on their absolute magnitude, but also on the interaction range. Crucially, even weak, but very long-ranged couplings can induce the topological phase transition. This elucidates the role of the interaction length as a new parameter controlling the emergence of topological phases and topological edge states. While here we investigated an abstract tight-binding model, our conclusions are directly relevant for  the variety of experimental platforms including the arrays of trapped ions~\cite{nevado2017topological, Dumitrescu2022, Edmunds2025, Katz2024}, arrays of dipolar scatterers~\cite{Newman2018,He2022} or lattices of optical waveguides coupled via extended photonic couplers~\cite{Vicencio2025}.

\section*{ACKNOWLEDGMENTS}

We acknowledge Maxim Mazanov and Konstantin Rodionenko for valuable discussions. This work was supported by the Russian Science Foundation grant No.~25-79-31027.

\section*{Data availability statement}

The data that support the findings of this study are available from the corresponding author upon request.

\bibliography{main}

\end{document}